# Identifying Trades Using Technical Analysis and ML/DL models

Aayush Shah[1], Mann Doshi[2], Meet Parekh[3], Nirmit Deliwala[4], Prof. Pramila M. Chawan[5]

1,2,3,4 B.Tech Student, Dept of Computer Engineering and IT, VJTI College, Mumbai, Maharashtra, India

5 Assistant Professor, Dept of Computer Engineering and IT, VJTI College, Mumbai, Maharashtra, India

**ABSTRACT:** The importance of predicting stock market prices cannot be overstated. It is a pivotal task for investors and financial institutions as it enables them to make informed investment decisions, manage risks, and ensure the stability of the financial system. Accurate stock market predictions can help investors maximize their returns and minimize their losses, while financial institutions can use this information to develop effective risk management policies. However, stock market prediction is a challenging task due to the complex nature of the stock market and the multitude of factors that can affect stock prices. As a result, advanced technologies such as deep learning are being increasingly utilized to analyze vast amounts of data and provide valuable insights into the behavior of the stock market. While deep learning has shown promise in accurately predicting stock prices, there is still much research to be done in this area.

## I. INTRODUCTION

The onerous and pioneering task of predicting stock market prices is an indispensable task for any informed investor. The non-stationary, non-linear, random, and noisy behavior of the market does not help make the task any less challenging. We can use modern technology and program models that can help predict potential future trends and patterns that emerge from observing several years of stock market data. These trends and patterns could help investors make informed decisions about stocks under their purview and maximize their profits. The fitful nature of the markets could intimidate even the most experienced and protean of investors and therefore it is momentous for the task to be automated to provide investors some sort of a structure to a mercurial market.

### A. Problem

Accurately predicting stock market prices is a grueling and painstaking task. Erudite investors make use of technical and fundamental analysis, which stem from certain technical indicators designed to evince which stocks to invest in so as to make a net profit. Even today, a large fraction of investors and traders manually design and adhere to a set of rules derived from the aforementioned technical & fundamental analysis. This process, however reliable it might be, is mind-numbing and transitory based on current market trends. Perhaps a more sophisticated and comprehensive approach would be to use deep learning models to automate the task and make it more generic in order to embrace a wider range of trends and patterns.

### B. Complexity

A plethora of models exist for predicting stock market prices. From all the available options, selecting the model that is the right fit for our use case is cumbersome. In addition to selecting the right model, we also have to pick the right hyperparameters for our model since they govern the learning rate of the model. This involves a lot of experimentation and trial and error done over a large dataset to arrive at the right blend. There are also times when a certain model is better suited for a certain period of interval of time or for a certain sector of the market. This makes the entire process of stock market prediction formidable and understandably, byzantine.

*C. Challenges*

The preliminary challenge was to access the publicly available APIs for obtaining accurate yet reliable stock market data, one needs to have a registered trading account, so that was a major drawback and also, the number of requests that could be sent to the server were limited. Since we were thinking of analyzing and making predictions for at least 500 stocks, we needed a thorough approach to this. Once this was dealt with, the major and the most intimidating challenge was to calculate and store the values of all the stock market indicators (approx 60) for all the stocks (approx 500). And the mere volume of the data suggested we use some of the more sophisticated processing methods that deal with the data.

## II. RELATED WORK

Various research studies have attempted to study stock market behavior. Deep learning has enabled us, from manually designing buying and selling rules using different technical and fundamental indicators to allowing the models to design complex rules for us using historical data. The model here acts as a black box. Deep Neural Networks, Support Vector Machines, Random Forests, and Decision Trees are some of the many models that can be used for this purpose.

Mabrouk et al. (2022) [1] discusses an Intraday Trading Strategy based on GRU and CNN for the Forex market. The authors used feature selection techniques to reduce the number of inputs from more than 150 variables to less than 30, keeping only the most important features. Their proposed model is composed of four layers, with the first layer taking a two-dimensional matrix as input. The authors evaluated their proposed model using real-world data from the Forex market and compared it with several baseline models. The results showed that their hybrid model had the best performance in terms of profit accuracy for predictions in all periods. It also reduced the number of transactions compared to the baseline models.

Agrawal et al. (2022) presents a deep learning model for predicting stock prices based on technical indicators. The authors propose a novel architecture that combines convolutional neural networks (CNNs) and long short-term memory (LSTM) networks to capture both spatial and temporal features of the input data. The proposed model is evaluated on real-world stock market data from the National Stock Exchange of India (NSE) and compared with traditional machine learning models such as support vector regression (SVR) and random forest regression (RFR). The results show that the proposed model outperforms these traditional models in terms of prediction accuracy, with an average mean absolute error (MAE) of 0.0125. The authors also conduct sensitivity analysis to investigate the impact of different hyperparameters on the performance of the model. Finally, the authors discuss some limitations and potential drawbacks of using deep learning models for stock prediction, such as overfitting, data quality issues, and interpretability challenges.

Hamoudi and Elseif (2021) [3] presents a novel approach for stock market prediction using 1-Dimensional Convolutional Neural Networks (CNN) and Long Short-Term Memory (LSTM) models. The goal of the project is to develop prediction models for high-frequency automated algorithmic trading using a data set of 130 anonymous intra-day market features and trade returns. The models are designed to predict the trades resulting in returns in the top ten percentile of all returns in the training set. The authors introduce two novelties in their approach. First, they frame the problem as a binary classification task rather than trying to predict the exact value of the return for a given trading opportunity. Second, they augment the feature matrix with a logical array to preserve information about missing features at each time step. The results show that their approach achieves positive returns with very low risk by identifying promising trading opportunities. The deep neural network models outperform traditional machine learning models such as Artificial Neural Networks (ANNs) and Support Vector Machines (SVMs).

Ghosh et al. (2021) [4] proposes a novel approach to forecast directional movements of stock prices for intraday trading by using Long Short-Term Memory (LSTM) and Random Forests (RF) models. The proposed approach involves using technical indicators such as moving averages, relative strength index (RSI), and Bollinger Bands as input features to train the models. The LSTM model is used to capture the temporal dependencies in the data, while the RF model is used to select the most relevant features and make final predictions. The authors evaluate the proposed approach on historical data of two major stock indices, namely the S&P 500 and NASDAQ. The results show that the proposed approach outperforms several benchmark models in terms of accuracy and profitability, as measured by the trading simulation results.

Faraz et al. (2020) [5] proposes a novel approach for predicting the closing price of the stock market using autoencoder long short-term memory (AE-LSTM) networks. The model uses an encoder-decoder-based architecture to extract invariant and abstract features from the raw dataset in an unsupervised way. Technical analysis techniques such as wavelet transformation and outlier exclusion are also employed to eliminate market noise and improve prediction accuracy. The z-score method is used for anomaly detection in the dataset, further improving the accuracy of the predictions. Overall, this paper presents a promising strategy for predicting stock prices using deep learning methods and technical analysis.

Tomar et al. (2020) [6] explores the use of deep learning techniques, specifically LSTM neural networks and their slim variants, for intraday stock trading. The study focuses on predicting the next-minute price movement of the US SPDR S&P 500 index using a dataset consisting of various technical indicator features. The paper outlines the methodology used in detail, including information about the dataset, features used, and feature selection. The results show that the slim LSTM2 model outperforms other models in terms of accuracy and F1 score. This study provides insights into how deep learning techniques can be applied to real-world data for intraday stock trading and could potentially be useful for traders or investors looking to make more informed decisions based on predictive modeling.

Miao (2020) [7] explores a deep learning approach for stock market prediction using a Long Short-Term Memory (LSTM) network. The LSTM model is trained on historical stock price data and tested on future stock prices to predict the trend of the stock market. The study tests different parameters such as the number of hidden layers, drop-out regularization, and batch size to determine their effect on result accuracy. The dataset used in this study includes historical stock prices of Amazon, Google, and Facebook from January 2010 to December 2017. The dataset is preprocessed by normalizing the data and dividing it into training and testing sets. The results show that the LSTM technique is widely applied in similar works and achieves promising results in predicting stock prices. The study finds that changing the number of neurons can influence the model's performance while changing the time step can also affect accuracy.

Paspanthong et al. (2019) [8] proposes the use of machine learning models to predict future price movements of SPDR S&P 500 trust and how investors can benefit from these predictions. The authors implemented multiple algorithms, including logistic regression, support vector machines

(SVM), Long-Short Term memory (LSTM), and Convolutional Neural Networks (CNN), to determine the trading action in the next minute. They used the predicted results from their models to generate the portfolio value over time and found that a support vector machine with a polynomial kernel performs the best among all of their models. The main goal of this paper is not only to assess the statistical performance of machine learning in forecasting future price movements but also effectively evaluate the results in terms of actual profits.

Agarwal et al. (2018) [9] presents a thesis on the development of a stock price prediction model using technical analysis and machine learning. The model uses a two-layer reasoning approach that employs domain knowledge from technical analysis to guide a second layer of reasoning based on machine learning. The model is supplemented by a money management strategy that uses the historical success of predictions made by the model to determine the amount of capital to invest in future predictions. The performance of the model is evaluated on stocks listed on the Oslo Stock Exchange, and it is found to outperform the Oslo Benchmark Index (OSEBX) successfully. This thesis provides valuable insights into the use of technical analysis and machine learning for predicting stock prices, which can be useful for investors and traders in making informed decisions.

Sezer et al. (2017) [10] proposes a model that optimizes technical analysis parameters using genetic algorithms (GA). The study uses Genetic Algorithm to optimize RSI parameters for different market conditions. The optimized feature values are then used as buy-sell trigger points for the deep neural network data set. The methodology of parameter optimization is an important aspect of the study, as it enhances the stock trading performance and provides a model that might be used as an alternative to Buy and Hold and other standard technical analysis models.

Kwon et al. (2007) [11] presents a hybrid neurogenetic approach for stock forecasting that combines a recurrent neural network and a genetic algorithm to optimize stock trading predictions. The system was tested on 36 companies in NYSE and NASDAQ for 13 years and showed notable improvement over the buy-and-hold strategy and the context-based ensemble. The authors also used a context-based ensemble method to dynamically adjust the predictions based on the test day's context. This paper provides valuable insights into the use of machine learning techniques for financial portfolio construction.

### III. PROPOSED SYSTEM

A. *Problem Statement*

To implement a model that analyzes the daily prices of all the stocks in an index and predicts trends thus helping in identifying profitable trades. This may help traders/investors to automate the task of shortlisting promising stocks and thus help them boost their profits.

B. *Proposed Methodology*

To predict buying points in stocks to gain profit, we use historical data of the current NIFTY 500 index of NSE. The data involves Open, High, Low, and Close (OHLC) prices of each trading day from March 2000 to March 2022 as well as the volume of equity shares traded.

We have used technical analysis exclusively as our trading method. Technical analysis involves statistical methods and chart patterns to determine price movement. We have used technical indicators and candlestick patterns that can be derived from the OHLC values. These statistical tools help in determining a strategy to buy stocks based on certain parameters and constraints.

The strategy is then implemented on the historic data and gives certain buying points and selling points. These buying and selling points are backtested to check if they are profitable or not. The output of the strategy is added to our data with values of -1 where the strategy doesn't give a buying signal, 0 where the strategy gives a buying signal but is proven to incur a loss when backtested and 1 where the strategy gives a buying signal and the trade is profitable when backtested.

Thus the problem has now been transformed into a classification problem where a machine learning model classifies the buying points of a strategy into profitable or lossy trade. All rows with a -1 value in the strategy attribute are removed and the remaining rows are further preprocessed and given as input. The machine learning model classifies the trade and gives output based on the trained weights thus helping a trader to strengthen his strategy further and maximize profits.

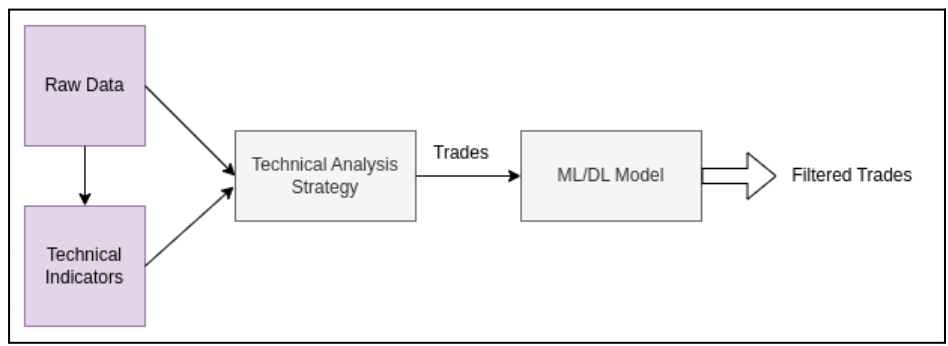

a. **Adding Features**

Apart from price features and volume, we have added technical indicators, candlestick patterns, and strategy features. Technical indicators are statistical values derived from the price or volume features that give an indication of the trend, momentum, or volatility of the stock. We derived the following technical indicators from the raw data so as to feed them to the models.

- RSI

The relative strength index or RSI is a momentum-based oscillating technical indicator. It indicates the strength of buying and selling for a given time period and also tells if the stock is overbought or oversold.

$$RSI = 100 - \left[ \frac{100}{1 + \frac{n_{up}}{n_{down}}} \right]$$

where:
$n_{up}$ = average of n-day up closes
$n_{down}$ = average of n-day down closes
(most analysts use 9 - 15 day RSI)

- MACD

Moving average convergence divergence(MACD) is a trend-based technical indicator that indicates potential trends using moving averages of the underlying security. It is derived from the difference between two variable-length exponential moving averages.Following are some of the technical indicators derived from the closing price of the underlying security to help train the model

> MACD line: 12-Period EMA – 26-Period EMA
>
> Signal line: 9-Period EMA
>
> Histogram: Difference between MACD line and signal line

- Bollinger Bands

  Bollinger Bands are volatility-based technical indicators. Bollinger bands indicate whether the stock has been bought or sold beyond certain values of standard deviations thus indicating the volatility. We have used Bollinger bands with 2 standard deviations.

  $$Upper\ Band = Moving\ Average + Constant \sqrt{\frac{\sum_{i=1}^{n}(y_i - Moving\ Average)^2}{n}}$$

  $$Lower\ Band = Moving\ Average - Constant \sqrt{\frac{\sum_{i=1}^{n}(y_i - Moving\ Average)^2}{n}}$$

- Simple Moving Average

  A simple moving average (SMA) is the rolling mean of the closing prices of the stock for a particular number of periods.

  $$SMA = \frac{A_1 + A_2 + ... + A_n}{n}$$

  **where:**
  $A_n$ = the price of an asset at period $n$
  $n$ = the number of total periods

- Exponential Moving Average

  The exponential moving average (EMA) is the weighted rolling mean of the closing prices of the stock for a particular number of periods. It applies more weight to the more recent prices.

  $$EMA_{Today} = Price\ Today \times \left(\frac{Smoothing}{1 + Days}\right) + EMA_{Yesterday}\left(1 - \left(\frac{Smoothing}{1 + Days}\right)\right)$$

- Directional Movement Index

  The Directional Movement Index (DMI) is an oscillation-type indicator that measures the trend of a security.

$$+DI = \left(\frac{\text{Smoothed } +DM}{\text{ATR}}\right) \times 100$$

$$-DI = \left(\frac{\text{Smoothed } -DM}{\text{ATR}}\right) \times 100$$

$$DX = \left(\frac{|+DI - -DI|}{|+DI + -DI|}\right) \times 100$$

where:

+DM (Directional Movement) = Current High − PH

PH = Previous high

−DM = Previous Low − Current Low

Smoothed +/−DM = $\sum_{t=1}^{14} DM - \left(\frac{\sum_{t=1}^{14} DM}{14}\right) + CDM$

CDM = Current DM

ATR = Average True Range

- Candlestick Patterns

    Candlestick patterns are technical patterns identified on a candlestick chart of a security. These patterns may give significant indications of a trend reversal or the start of a trend. We have used 13 candlestick patterns in our data with a boolean data type where a true value indicates the formation of a candlestick pattern on that particular day for a given stock.

**b. Data Preprocessing**

**Data Normalization:** We use data normalization so as to transform the data such that they have no/similar dimensions. Variables that are bigger in magnitude tend to steer the model in their direction which is not good for performance. Data normalization helps prevent such behavior by ensuring each variable has similar weights/importance. We have used min-max normalization for data normalization because it ensures that all features will have the same scale.

*C. Technical Analysis Strategy*

SMAs and Bollinger Bands are some examples of mathematically-derived technical indicators that traders use. They help them analyze past price trends and patterns and predict future trends to an extent. Fundamental investors rely more on financial reports of the company, market research, and business profitability to identify a worthy investment. On the other hand, technical traders rely on price charts and various other statistical tools to predict price trends and identify trading opportunities.

Technical Indicators help identify trading opportunities. For example, a moving average crossover often signifies a change in an ongoing trend. In this instance, applying a moving average indicator to a stock's price chart can help identifying an area where an ongoing price may change, thus identifying a trading opportunity.

a. **Breakout Resistance Strategy**

    Resistance level is a price at which selling pressure exceeds the buying momentum and acts as a barrier to the stock trend. Stock prices tend to decline at this level due to psychological selling prices set by traders. A breakout occurs when the stock price breaks above this resistance level and creates new highs. The breakout resistance strategy gives buying signals during this period. The resistance level is determined by the previous maxima

created by the stock chart pattern. Whenever the stock breaks above the resistance level, a buy signal is generated. The selling point is determined by a stop loss and a trailing loss. Stop loss is a certain price point below the buying price. If the stock price goes below the stop loss, a selling signal is triggered where the stock is sold at a loss. Trailing loss is similar to stop loss except it moves dynamically as the stock price makes new highs.

D. *Model*
   a. **Neural Network Classification**

Deep neural networks (DNNs) are a type of machine learning algorithm inspired by the structure and function of the human brain. They consist of multiple layers of interconnected nodes, or neurons, that process input data and progressively extract higher-level features for classification or other tasks. DNNs are particularly useful for classification tasks because they can automatically learn complex decision boundaries from large and high-dimensional datasets. In other words, they can classify objects or events into different categories based on patterns and relationships in the data, without being explicitly programmed to do so.

As input to the model, we have provided trades that are shortlisted by the technical trading strategy. Each row, signifying the buying point of the trade, contains 51 attributes. The model outputs a single value for each row. This single output is the probability of the trade being profitable.

For the loss function, we have used binary cross-entropy. We use RMSProp as the optimizer. The model has one hidden layer with 9 hidden units.

```
Model: "sequential"
_________________________________________________________________
 Layer (type)                Output Shape              Param #
=================================================================
 dense (Dense)               (None, 20)                1040

 dense_1 (Dense)             (None, 10)                210

 dense_2 (Dense)             (None, 1)                 11

=================================================================
Total params: 1,261
Trainable params: 1,261
Non-trainable params: 0
```

   b. **XGBoost**

XGBoost (short for Extreme Gradient Boosting) is a type of machine learning algorithm that uses decision trees to model and predict outcomes. It is an ensemble method, which means it combines the predictions of multiple weaker models (decision trees) to make a more accurate final prediction.

XGBoost works by iteratively training decision trees on the residuals (errors) of the previous trees, gradually improving the overall prediction accuracy. It also uses regularization techniques to prevent overfitting, which can occur when a model is too complex and fits the training data too closely.

We use XGBoost to classify the trades given by the technical analysis strategy. We use the following hyperparameters:

**max_depth** = 5, **learning_rate** = 0.1, **n_estimators** = 500

c. **Decision Tree**

Decision Tree, in various papers referred to and in the past have shown constructive results in binary classification problems. Also analyzing the decision tree leads to better feature selection for other methods of classification too. We have leveraged the decision tree to classify buy signals received from our strategy.

The input consisted of 35 features ranging from normalized close values, binary data of different types of Candle sticks, and encodings of the Stock's company name and its respective industry. We used Label Encoding to encode Company and Industry names here. We explored hash encodings too but for the Decision Tree label, encoding was a better choice since hash encodings led to an output of 6 bits which is a better input for a Neural network and not a decision tree. Gini Impurity was used in the Decision tree since it is ideal for binary classification. We used Graphviz for visualizing the resulting Decision tree.

A snapshot of the resulting Decision Tree is shown below.

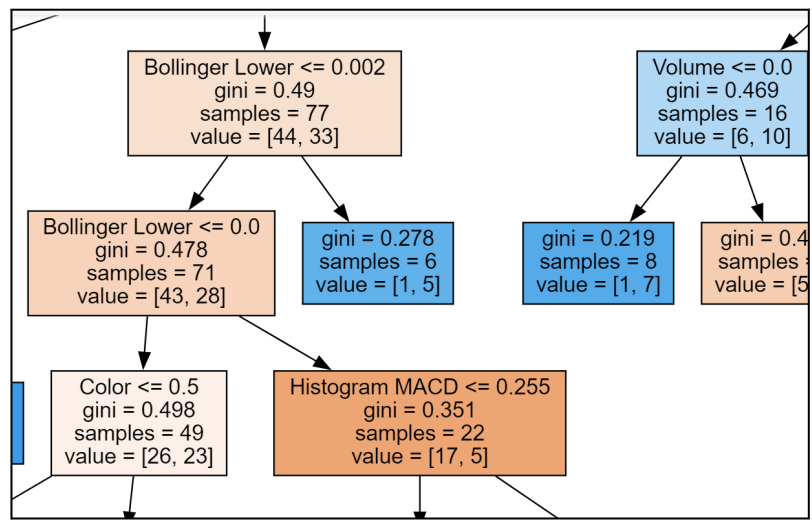

The resulting Decision Trees were of large size, leading to which we set the max depth to 6.

IV. RESULTS AND DISCUSSION

The proposed methodology of using logistic regression to filter further the trades shortlisted by the technical trading strategy provided significant improvement in identifying profitable trades. To understand how the proposed method would perform in a real-world scenario, we test the model using the rolling cross-validation approach. For eg - We train the model on data from 2000 to 2011 and test the model on data from 2011 to 2012, and similarly test it on further years. This also helps us understand how the model will perform in constantly changing market conditions.

■ **Precision/Recall table**

**Using only Strategy:**

|  | Precision |
|---|---|
| **2011-12** | 0.3559 |
| **2012-13** | 0.3572 |
| **2013-14** | 0.4448 |
| **2014-15** | 0.4484 |
| **2015-16** | 0.3263 |
| **2016-17** | 0.4264 |
| **2017-18** | 0.4020 |
| **2018-19** | 0.3480 |
| **2019-20** | 0.3357 |
| **2020-21** | 0.4570 |
| **2021-22** | 0.3858 |

**Using Decision Tree:**

|  | Precision | Recall |
|---|---|---|
| **2018-22** | 0.39 | 0.32 |

Since the performance of the Decision Tree was not promising we discontinued it further and have not performed a year-wise analysis for the same.

**Using Neural Network**:

|  | Precision | Recall |
|---|---|---|
| **2011-12** | 0.3778 | 0.1228 |
| **2012-13** | 0.4028 | 0.1448 |
| **2013-14** | 0.4941 | 0.1478 |
| **2014-15** | 0.4513 | 0.0414 |
| **2015-16** | 0.2807 | 0.0224 |
| **2016-17** | 0.4051 | 0.0247 |
| **2017-18** | 0.3654 | 0.0690 |
| **2018-19** | 0.3803 | 0.0999 |
| **2019-20** | 0.4008 | 0.1257 |
| **2020-21** | 0.5069 | 0.1494 |
| **2021-22** | 0.4503 | 0.0598 |

**Using XGBoost:**

|  | **Precision** | **Recall** |
|---|---|---|
| **2011-12** | 0.3889 | 0.3035 |
| **2012-13** | 0.4072 | 0.2427 |
| **2013-14** | 0.5050 | 0.2225 |
| **2014-15** | 0.5266 | 0.1852 |
| **2015-16** | 0.3667 | 0.1695 |
| **2016-17** | 0.4665 | 0.1507 |
| **2017-18** | 0.4492 | 0.1325 |
| **2018-19** | 0.4052 | 0.1392 |
| **2019-20** | 0.3831 | 0.1198 |
| **2020-21** | 0.4915 | 0.1179 |
| **2021-22** | 0.4031 | 0.0677 |

    We can see from the above tables that the model does a great job in filtering the trades given by the technical analysis strategy. But using precision and recall as a metric will not suffice as we do not know profit per trade. It does not do any good if the model has high precision but only identifies many trades with very small profits and fewer trades with huge losses. On the other hand, it is okay to have a model with lower precision but that identifies fewer trades with huge profits and many trades with small profits.

To have a better idea regarding the performance of the model we can compare the models by calculating the average profit per trade for the trades that are predicted to be profitable by the model.

■  **Average Profit per Trade Table**

|         | **Only Strategy** | **Neural Network** | **XGBoost** |
|---------|-------------------|--------------------|-------------|
| **2011-12** | 2.3   | 3.067 | 2.865 |
| **2012-13** | 2.39  | 3.424 | 3.271 |
| **2013-14** | 3.977 | 4.63  | 4.828 |
| **2014-15** | 4.538 | 5.859 | 6.758 |
| **2015-16** | 1.844 | 1.703 | 2.288 |
| **2016-17** | 4.024 | 4.806 | 5.056 |
| **2017-18** | 3.252 | 2.999 | 4.331 |
| **2018-19** | 1.73  | 3.176 | 3.259 |
| **2019-20** | 2.119 | 4.526 | 4.592 |
| **2020-21** | 5.185 | 7.752 | 8.771 |

V.  CONCLUSION

By using the initial strategy developed by us we were able to achieve an average of 3.1359 % profit per trade each year since 2011 with an average precision of 0.3897. Using Neural Network Architecture, we improved it to 3.9883 % ( a 27% improvement from the strategy-only method) with an average precision of 0.4105 ( a 5% improvement). Further XGBoost logged 4.6019 % profit per trade each year since 2011 (a 46% improvement) with an average precision of 0.4357(an 11.8% improvement). Thus, according to the proposed methodologies, applying DL/ML techniques on top of trades given by technical analysis strategies increases average precision and per-trade profitability significantly.

VI. FUTURE WORK

In the future, we hope to experiment with more models. We also wish to modify the models such that they take into account the magnitude of profit/loss with the help of multi-class classification or by making the model predict the expected profit instead of just making it predict whether the trade will be profitable or not. Furthermore, we also wish to use a similar approach to a finer/broader timeline i.e. to hourly/weekly charts.

**BIOGRAPHY**

**Aayush Shah**, B. Tech Student, Dept. of Computer Engineering and IT, VJTI College, Mumbai, Maharashtra, India.

**Mann Doshi**, B. Tech Student, Dept. of Computer Engineering and IT, VJTI College, Mumbai, Maharashtra, India

**Meet Parekh**, B. Tech Student, Dept. of Computer Engineering and IT, VJTI College, Mumbai, Maharashtra, India.

**Nirmit Deliwala**, B. Tech Student, Dept. of Computer Engineering and IT, VJTI College, Mumbai, Maharashtra, India.

**Prof. Pramila M. Chawan** is working as an Associate Professor in the Computer Engineering Department of VJTI, Mumbai. She has done her B.E. (Computer Engineering) and M.E. (Computer Engineering) from VJTI College of Engineering, Mumbai University. She has 28 years of teaching experience and has guided 80+ M. Tech. projects and 100+ B. Tech. projects. She has published

134 papers in International Journals and 20 papers in National/International Conferences/ Symposiums. She has worked as an Organizing Committee member for 21 International Conferences and 5 AICTE/MHRD-sponsored Workshops/STTPs/FDPs. She has participated in 14 National/International Conferences. She has worked as NBA Coordinator of the Computer Engineering Department of VJTI for 5 years. She had written a proposal under TEQIP-I in June 2004 for 'Creating Central Computing Facility at VJTI'. Rs. Eight Crores were sanctioned by the World Bank under TEQIP-I on this proposal. The Central Computing Facility was set up at VJTI through this fund which has played a key role in improving the teaching-learning process at VJTI.